\documentstyle[twocolumn,prc,aps,psfig]{revtex}

      \newcommand{\pathnow}{}
\begin{document}\hbadness=10000
\twocolumn[\hsize\textwidth\columnwidth\hsize\csname %
@twocolumnfalse\endcsname
\title{Strange hadron resonances as a signature of freeze-out dynamics}
\author{Giorgio Torrieri and Johann Rafelski}
\address{
Department of Physics, University of Arizona, Tucson, AZ 85721}
\date{March 14, 2001}
\maketitle
\begin{abstract}
\noindent
We study the production and the observability of 
$\Lambda^{*}(1520)$, $K^{*0}(892)$,  and $\Sigma^{*}(1385)$, 
strange hadron resonances as function of the freeze-out 
conditions  within the statistical model of hadron production.  
We obtain an estimate of how many  decay products 
are rescattered in evolution towards thermal freeze-out 
following chemical freeze-out, 
and find that the resonance decay signal is
strong enough to be detected.  We show how a combined 
analysis of at least two resonances can be used 
to understand the chemical
freeze-out temperature, and the time between chemical
 and thermal freeze-outs.\\

\end{abstract}
\vspace{+0.5cm}
]
\begin{narrowtext}
Hadronic particle signatures of the formation of QGP phase 
in relativistic heavy ion collisions are most sensitive when 
the dense hadron matter fireball breakup is sudden \cite{Raf00,Let00}.
But  final state particles could also emerge remembering relatively
little about their primordial source, having been subject
to rescattering in purely hadronic gas phase \cite{PBM99,CERN}. 
Which reaction picture applies can have decisive 
influence on our understanding 
of the underlying physics, and thus we propose here a systematic
method to  experimentally make a distinction.

To address this question we consider
strange hadron resonance abundances. 
At this time  $\Lambda(1520)$ has
been observed in heavy ion reactions at SPS energies 
 \cite{NA49Res,NA49Res2} following a suggestion 
that such a measurement was possible \cite{Bec98}.
SPS \cite{NA49Res2} and RHIC experiments \cite{STARkstar}
report measurement of the $\overline{K^{*0}}(892)$ signal, and
RHIC has already measured both the $K^{*0}$ and the
$\overline{K^{*0}}$.  In the SPS case,
the $\Lambda(1520)$ abundance yield is about 2.5 times smaller
than expectations based on the yield 
extrapolated from nucleon-nucleon reactions. This is 
to be compared with the  enhancement by factor 2.5 of 
$\Lambda$-production in the same reaction in terms of the 
same comparison.  

A possible explanation for
this effective suppression by a factor 5, or more, is that the
decay products ($\pi,\Lambda$) produced at a high 
chemical freeze-out temperature $T\simeq 175$\,MeV 
 have rescattered and thus their momenta
did not allow  to reconstruct this state in an invariant mass analysis.
However, the observation of a strong $K^{*0}$-yield signal  
contradicts this point  of view, since the $K^{*0}(892)$ decays faster
($\Gamma_{K^{*0}}=50.8\pm0.9\,$ MeV  $>$ 
$\Gamma_{\Lambda^*}=15.6\pm 1\,$ MeV). Therefore, 
$K^{*0}(892)$ should be even more  suppressed: 
a back of envelope calculation based 
on exponential population attenuation
suggests that if the observable yield of  $\Lambda(1520)$ is reduced 
by factor 5, the observable 
yield of $K^{*0}(892)$  should be suppressed by a factor 10.

Another explanation is that the chemical freeze-out temperature
which governs the production of these resonances in a thermal 
model is considerably lower, and hence rescattering of decay 
products is less significant. This is 
a point of view arising from a recent analysis of the 
hadronization process \cite{Raf00}.

We explore the production, and the suppression of 
the observability of
these resonances($K^{*0}(892),\,\Lambda(1520)$), and   
also explore  the (more difficult) 
measurement of the $\Sigma^{*}(1385)$ state as a further 
way of distinguishing
between  reaction scenarios. We will show in a quantitative 
analysis how these
measurements can constrain both the chemical
freeze-out temperature and the lifetime of the hadron interacting
phase evolving  between the chemical and the thermal freeze-out conditions.

The $\Sigma^*(1385)$ is expected to be produced in a thermal
model more abundantly 
than $\Lambda^*(1520)$ in a hadronic fireball due to it's high 
degeneracy factor and smaller mass. 
Because of its 3 times shorter 
lifetime ($\Gamma_{\Sigma^*}=36$--$39\,$ MeV  $>$ 
$\Gamma_{\Lambda^*}=15.6\pm 1\,$ MeV), the $\Sigma^*$ signal 
is  more strongly influenced by final state interactions than that
of  $\Lambda^*(1520)$.  Like for $K^{*0}(892)$,  one would naively 
expect that the observable 
yield of $\Sigma^*$  should be suppressed by a factor 10.


We will first  show that, in a fireball hadronizing according
to the statistical model, the proportion of $\Lambda$ s  and Kaons 
resulting from decays of the main resonances can be
noticeably dependent on chemical freeze-out temperature.
Hadrons produced directly from a medium at temperature ($T$) much lower
than the particle mass (as in all cases considered here) 
fill the available 
statistical phase space which has the relativistic Boltzmann 
distribution shape:
\begin{equation}
\label{direct}
\frac{d^2 N}{dm_{T}^{2} dY} \propto 
g \prod^{n}_{i=1} \lambda_{i} \gamma_{i}  m_{T} \cosh(y) e^{-E/T} 
\end{equation}

Here $g$ is the statistical degeneracy, $\lambda_{i}$ and
$\gamma_{i}$ are the fugacity and equilibrium parameters
of each valence quarks, and E is the energy. When
 the fireball is expanding at a relativistic speed,
equation \ref{direct} 
describes the energy distribution of an element of the
fireball in a reference frame at rest with respect to the expansion
(flow). However, in this paper we evaluate  ratios of particles 
with similar masses, and interaction modes,  often 
considered in full phase space. For this reasons to a good 
approximation,  the flow effects largely cancel out. Similarly, 
for ratios of particles with the same valence quark composition,
such as $\Sigma^*/\Lambda,\,\Lambda(1520)/\Lambda$ and,
in the limit of $\lambda_{u}=\lambda_{d}$ (assumed
hence forward), 
 $K^{*0}(892)/K^-$
($=\overline{K^{*0}}/K^+$) 
the chemical factors ($\lambda$'s and $\gamma$'s) cancel out between
the two states compared.


We now include in the yields the descendents
arising from resonance decays. 
In a decay of the type $R \rightarrow 1+2$
where 
a particle R (mass M, transverse mass $M_T$, rapidity Y)
into a particle 1 (mass m, transverse mass $m_T$, rapidity y)
and  2 (mass $m_2$), the distribution
$d^2 N_{1}/ dm_{T}^{2} dy$ can be obtained in terms of 
$d^2 N_{R}/ dM_{T}^{2} dY$ using the following 
transformation :
\begin{eqnarray}
\label{reso}
\frac{dN_1}{d {m^2_{T}} d y } =
\frac{g_{r} b}{4 \pi p^{*}}
\int_{Y-}^{Y+} dY
\int_{M_{T}-}^{M_{T}+} dM_{T}^{2} J 
\frac{d^2 N_{R}}{dM_{T}^{2} dY}  \nonumber
\\
\noindent J=\frac{M}{\sqrt{P_{T}^2 p_{T}^2 -(M E^{*} - M_{T} m_{T} \cosh \Delta Y)^2 }}
\end{eqnarray}
Here $\Delta Y=Y-y$, $\sqrt{s}$ is the combined invariant mass of all the 
decay products
except 1, and $E^{*}=\frac{1}{2M}(M^2-m^2-m_2^2)$, $p^{*}=\sqrt{E^{*2}-m^2}$ 
are the energy and momentum of decay
particle 1 in the resonances rest frame.
$g_{r}$ accounts for a resonances degeneracy, 
while $b$ is the branching ratio
of the considered decay channel.

A full justification for Eq.\,(\ref{reso}) is found in 
Refs. \cite{resonances,resonances2}, where it is obtained 
by transforming the resonance distribution  from it's
rest frame into the fireballs rest frame, 
and integrating over all the
kinematically allowed values of rapidity and transverse mass.
$J$ is the Jacobian of this transformation,
with integration  limits:
\[\ Y_{\pm}=y \pm \sinh^{-1}(\frac{p^{*}}{m_{T}}) \]
\[\ M_{T}^{\pm}=M 
\frac{E^{*} m_{T} \cosh(\Delta Y) \pm p_{T} \sqrt{p^{*2}-m_{T}^{2} \sinh^{2} (\Delta Y)}}
{m_{T}^{2} \sinh^{2} (\Delta Y)+m^{2}}\]


Table \ref{processes} summarizes the dominant decay processes 
considered in our analysis and their parameters (Clebsh-Gordon 
coefficients have been used to estimate decays such as
($N^{*0} \rightarrow N^+ \pi-)/(N^{*0} \rightarrow N^0 \pi^0)$).

\begin{table}[h]
\centering
\caption{$\Lambda$ production mechanisms}
\begin{tabular}{|c|c|c|c|c|} 
g & Reaction & $p^{*}$  & branching & visible? \\ 
&& MeV & ratio & \\ 
\hline
$\approx 4$ & $\Sigma^{* 0}(1385) \rightarrow \Sigma^{0} \pi^{ 0}$ 
& 127 & $ \approx 4 \%$   & No
\\  \hline  
8 & $\Sigma^{* \pm}(1385) \rightarrow \Lambda \pi^{\pm}$ &
208 & $88 \%$  & Yes  
\\  \hline  
4 & $\Sigma^{* 0}(1385) \rightarrow \Lambda \pi^{0}$ &
208 & $88 \%$   & No
\\  \hline  
2 & $\Sigma^{0} \rightarrow \Lambda \gamma$ & 74  & $100 \%$ & No
\\  \hline  
4 & $\Lambda(1520) \rightarrow N \overline{K}$ & 244 & $45 \%$ & Yes
\\ \hline \hline
3 & $K^{*\,0}(892) \rightarrow K^+ \pi^-$ & 291  & $67 \%$ & Yes \\ \hline
3 & $K_{1}(1270) \rightarrow K \rho$ & 76 & $41 \%$ & No \\ \hline
3 & $K_{1}(1270) \rightarrow K^{*0}(892) \pi$ & 301 & $16 \%$ & No \\ \hline
6 & $K^{*0}(1400) \rightarrow K^{*0}(892) \pi$ & 301 & $100 \%$ & No \\ \hline
6 & $K^{*0}(1400) \rightarrow K^{*0}(892) \pi$ & 301 & $100 \%$ & No \\ \hline
5 & $K^{*0}(1430) \rightarrow K^{*0}(892) \pi,\pi \pi$ & $\approx 400$ & $40 \%$ & No \\ \hline
5 & $K^{*0}(1430) \rightarrow K  \pi$   & 622 & $50 \%$ & No \\ \hline
1 & $K^{*0}(1430) \rightarrow K  \pi$   & 622 & $100 \%$ & No \\ \hline
\end{tabular}
\label{processes}
\end{table}

In Fig. \ref{prodratios} we show the relative thermal  production ratios
at chemical freeze-out  over the entire spectrum
of rapidity and $m_{T}$  (solid lines) as well a central rapidity range 
defined by the $y-m_{T}$ region covered by  the
WA97 experiment ($|y| \le 0.5$ in
the center of mass frame) (dashed lines). 
The opportunity to measure the chemical freeze-out temperature by
a measurement of the relative resonance  yields is apparent.

For example,  at the lowest current estimates ($T\simeq 100$ MeV)
of the final break up temperature in 158$A$ GeV SPS collisions 
$33 \%$ of $\Lambda$ s are actually primary $\Sigma^{*}$ s, the percentage
rises to slightly more than $50 \%$ if chemical freeze-out
occurs at $T= 190$ MeV. We recall that the CERN WA97 collaboration 
has measured hyperon
yields to considerably greater precision than required
to distinguish between these limiting cases
 \cite{yields}, and the same is certainly expected of RHIC experiments
such as STAR.

Considering the practical
feasibility of such a measurement it should be noted that
the decay $\Sigma^{*} \rightarrow \Lambda \pi$ would appear
as a $\Lambda$-$\pi$ pair arriving from the central fireball.
Experiments usually reconstruct $\Xi \rightarrow \Lambda \pi$ decays by
finding the invariant mass of $\Lambda \pi$ pairs with a common
origin outside the collision, so a
$\Sigma^{*}$ event would normally be discarded as a pair of unrelated particles
coming directly from the fireball. An invariant mass
analysis with resolution better than $\Sigma^{*}(1385)$
natural width (35\,MMeV) would be needed to distinguish 
it from from  $\Xi(1321)$ and minimize the combinatoric background.

\begin{figure}[tb]
\begin{center}
\vspace*{-2cm}
\hspace*{.1cm}
\psfig{width=10cm,clip=1,figure=\pathnow 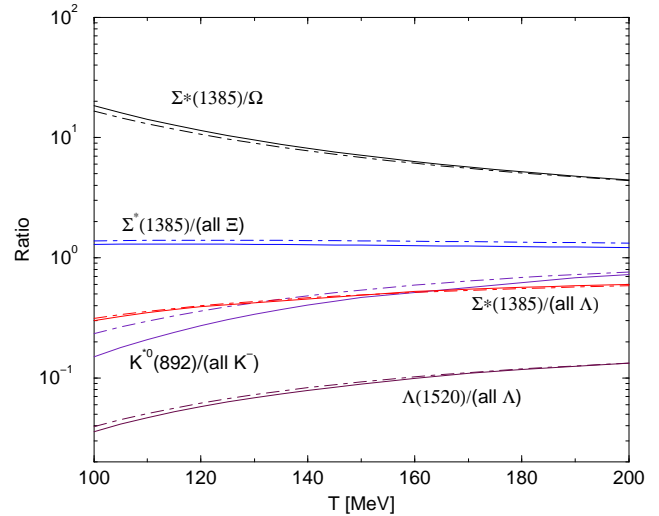   }
\vspace*{-0.21cm}
\caption{ 
Temperature dependence of ratios of $\Sigma^{*}$, $K^{*\,0}$
and $\Lambda(1520)$ to the total number of observed $K^+$
$\Lambda$ s,$\Xi$ s and $\Omega$ s. Branching ratios are included.
Dashed lines show the result for a measurement
 at central rapidity $\Delta y=\pm0.5$.
 \label{prodratios}}
\end{center}
\end{figure}
\vskip -0.5cm

The momentum of each particle in the $\Sigma^{*} \rightarrow \Lambda \pi$
pairs CoM (center of momentum) frame
would be $\approx$ 208 MeV/c, very close to the 211 MeV/c of the 
$ \Omega \rightarrow \Lambda K $ decay.
As Fig. \ref{prodratios} shows (top two solid/dashed lines), 
the potentially observable  (i.e. produced) $\Sigma^*$ s
should be more abundant than these two hyperons ( $\Xi, \Omega$) at
all considered hadronization temperatures.
 The chemical potentials
required to compute the two ratios 
$\Sigma^{*}/\Omega, \Sigma^{*}/\Xi$ in 
 Fig. \ref{prodratios}  were taken from \cite{Let00}.

The ratios of observed particles, however, 
can be considerably different from production ratios, since 
if the decay products rescatter before 
detection their identification by reconstructing 
their invariant mass will generally not be possible.
While the lifetime of the $\Xi$ and $\Omega$ are large enough to ensure that only
a negligible proportion of particles decay near enough to the fireball
for rescattering to be a possibility, the lifetime
of  $K^{*0}, \Sigma^{*}$  and even $\Lambda(1520)$ is within 
the same order of magnitude of the fireballs dimensions i.e. 
$2R/c\simeq 1/\Gamma$.
For this reason, a considerable number of decay products will undergo 
rescattering, and the estimation of this percentage is required before
any meaningful parameters are extracted from the data. 

Note in Fig.\,\ref{prodratios} that the 
relative $\Sigma^{*}/\Xi$ signal is remarkably independent
(within  5 \%) of Temperature.    This feature arises because 
the $\Xi^*(1530)$ contribution cancels nearly exactly the thermal
suppression of the $\Xi$ originating in the $\Xi-\Sigma^{*}$ mass difference.
This effect  could be used for a direct estimate of the
$\Sigma^{*}$ lost through rescattering,
 even without knowledge of the freeze-out 
temperature, should  the chemical
parameters ($\lambda_{i}$ and $\gamma_{i}$) be
known. A simple test of hadronization model 
consists in measurement of the ratio $\Sigma^{*}/\Xi$.
If it is significantly smaller than unity,
we should expect a re-equilibration mechanism to be present. Otherwise 
sudden hadronization applies. 

We can go further in the use of  the suppression
of the considered resonances  as a tool capable
of estimating conditions at particle freeze-out.  To do this we need
to estimate the effect rescattering would have on the resonances signal.
We formulate a simple model based on the width of the resonances in question 
and the decay products reactions  within an
expanding fireball of nuclear matter.
While more realistic models, possibly involving event generators
are needed for precise quantitative analysis, 
the model described below allows us  to make a qualitative prediction
and to verify the feasibility of the $\Sigma^{*}$ measurement.

We consider the decay of a generic resonance $N^{*}$
\begin{equation}
\label{decay}
N^{*} \rightarrow p_{1} p_{2}\,,
\end{equation}
in a gas of pions and nucleons.
For the invariant mass of $N^{*}$ to be undetectable, it is assumed here
that it is sufficient that either of the generic decay products 
$p_{1}$ or $p_{2}$ undergo one interaction.
This interaction depends on the cross-section of the decay product
with each particle in the fireball (pions, nucleons
and antinucleons), the speed of each decay product
relative to a typical fireball particle
and the pion and nucleon density in the fireball (decreasing with time for
a fireball constantly expanding with flow v).

This reaction rate  is
\begin{equation}
\label{scatter}
P_{1} = 
(\sigma_{1 \pi} \rho_{\pi}+\sigma_{1 N} \rho_{N} +\sigma_{1 \overline{N}} \rho_{\overline{N}}) 
(\frac{R}{R+v t})^3 \langle v  \rangle\,,
\end{equation}
and analogously for the second decay product `2'.
Here, $\sigma$'s are the (energy averaged)
interaction cross sections, the $\rho$'s are the densities,
$R$ is the fireball radius at chemical hadronization, 
$\langle v \rangle$ the velocity
factor and v is the space-average of the flow velocity.

To a good approximation, the thermal velocities in the fireball average out and 
$\langle v \gamma \rangle_{1,2}=p^{*}/m_{1,2}$, here $p^{*}$ is as 
defined in Eq.\,(\ref{reso}).
In addition to scattering and decay we also in 
principle can allow that $\Sigma^{*}$ and  $\pi$ each 
have  a  probability of escaping the fireball. We 
have studied this effect and  in this study 
of the measureability of the lifetime of the interacting
phase,  such an escape probability is not relevant.
The population equation describing the scattering loss abundance
($N_i$) are, therefore:
\begin{eqnarray} \nonumber  
 \frac{d N_{i}}{d t} &=& \frac{1}{\tau} N_{N^{*}} -N_{i} P_{i} \,, \quad i=1,2 \\[0.4cm] 
 \frac{d N_{N^*}}{d t} &=& -\frac{1}{\tau} N_{N^{*}}   \,,    
 \label{model}
\end{eqnarray}

The required nucleon density $\rho_{N}$ is obtained  in the 
a relativistic Boltzmann approximation: 
\begin{equation}\label{relboltz}
\rho_{N}=\frac{g}{(2 \pi \hbar c)^3} 4 \pi m^2 (\lambda_q \gamma_q)^3 T K_2 (\frac{m}{T}) \,,
\end{equation}
where the $K_{2} (x)$ is the 2nd modified Bessel function.
We consider the nucleons to have a mass of $\simeq$1 GeV, and a degeneracy of
6, to take the p,n and the thermally suppressed but higher degeneracy
$\Delta$ contributions into account in a first approximation.
Considering the CERN/SPS environment we calibrate 
the fireball size $R$ to obtain the 
baryon multiplicity to be $\approx$ 360 within the volume, and we take
$R=8$\,fm$\cdot 145/T$[MeV].

The temperature scaling of $R$ is chosen to assure that with 
the pion density computed in the massless particle limit, leading to
\begin{equation}\label{pidens}
\rho_{\pi}= \frac{A \pi^2}{90} T^3\,,
\end{equation}
the entropy of the expanding system is conserved. 
The effective degeneracy $A$ is
chosen to fix the entropy per baryon
to be equal to the observed number of $\simeq$40.
We believe that this qualitative model of baryon and meson 
density has the right 
physical magnitude and scales correctly with 
variation of freeze-out temperature.

To estimate the average cross sections needed in Eq.\,(\ref{scatter})
we recalled that the NA49 collaboration  using
the UrQMD event generator \cite{urqmd},  has found that 
$\approx 50 \%$ of $\Lambda(1520)$ undergo rescattering after a time of 80 fm/c.
In order to reproduce this result at  $T\approx 160$ MeV, 
we took scattering cross sections  shown in the top portion of 
table \ref{parameters}, based on Refs. \cite{crossections,Mar76}. 
It turns out that the cross section on nucleons and antinucleons are
sufficient to make the fireball  essentially opaque to 
one of the resonance decay products. 

\begin{figure}[tb]
\begin{center}
\vspace*{-1.8cm}
\psfig{width=9.5cm,clip=1,figure=\pathnow 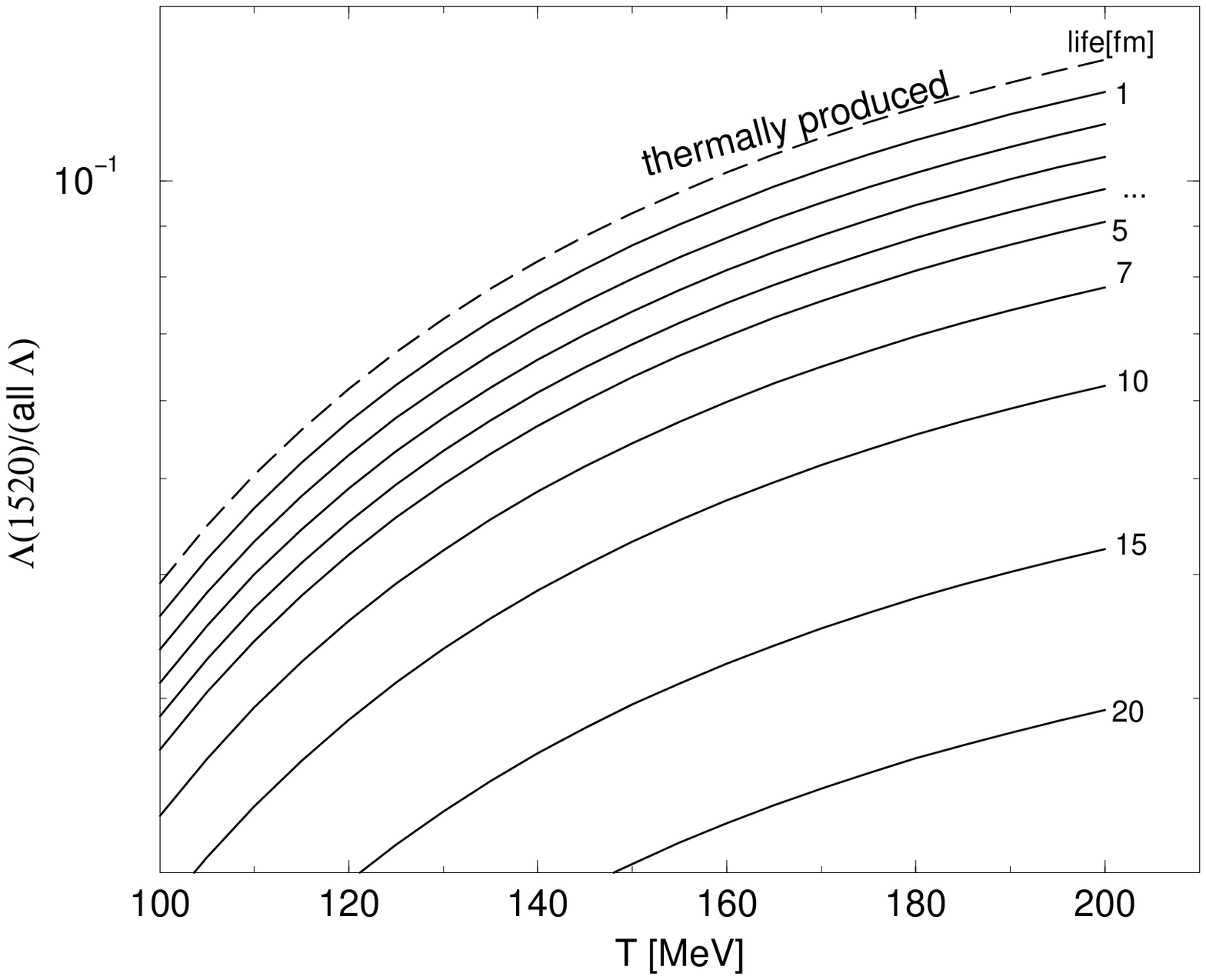 }
\vspace*{-1.5cm}
\psfig{width=9.5cm,clip=1,figure=\pathnow 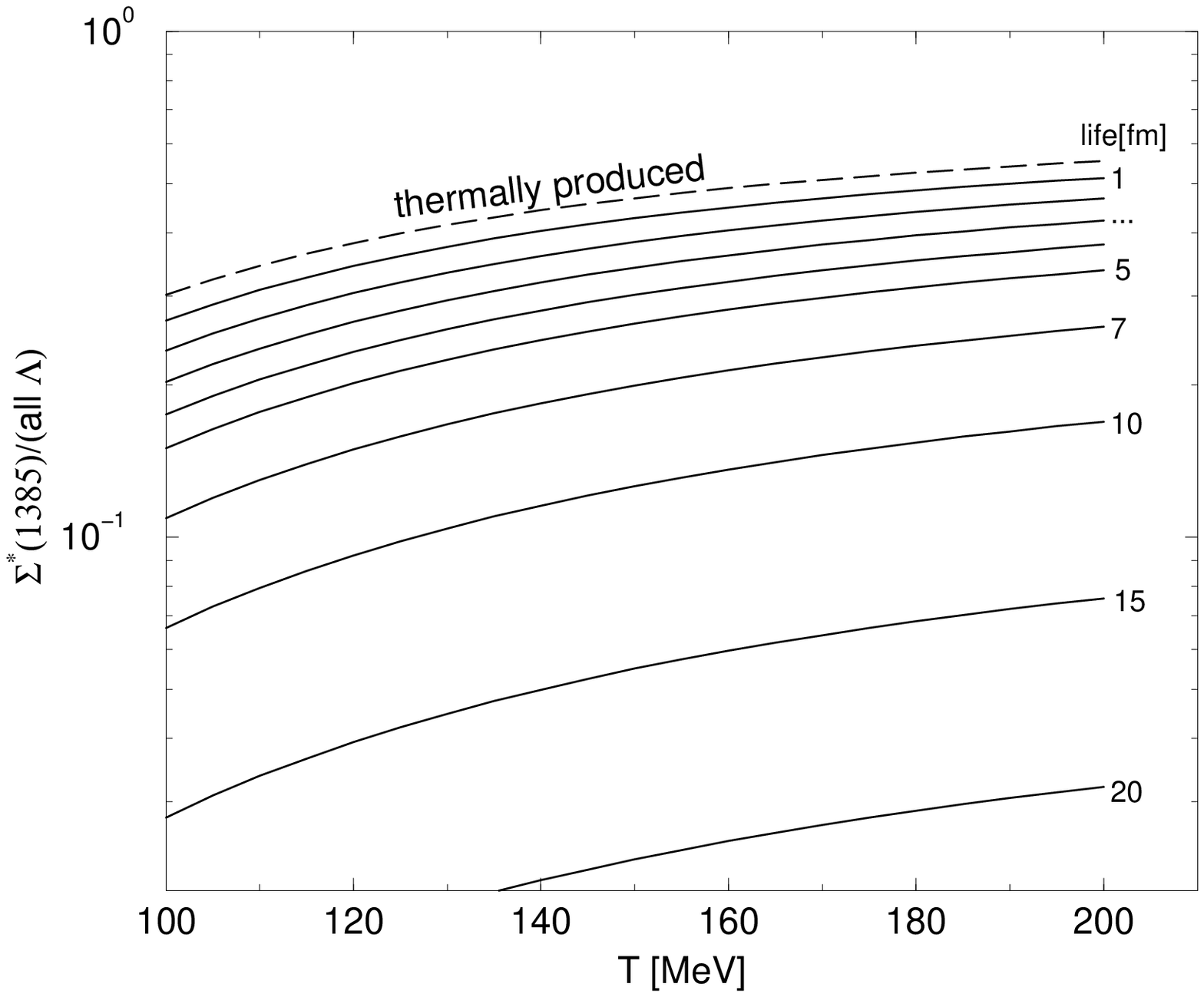}
\vspace*{-1.5cm}
\psfig{width=9.5cm,clip=1,figure=\pathnow 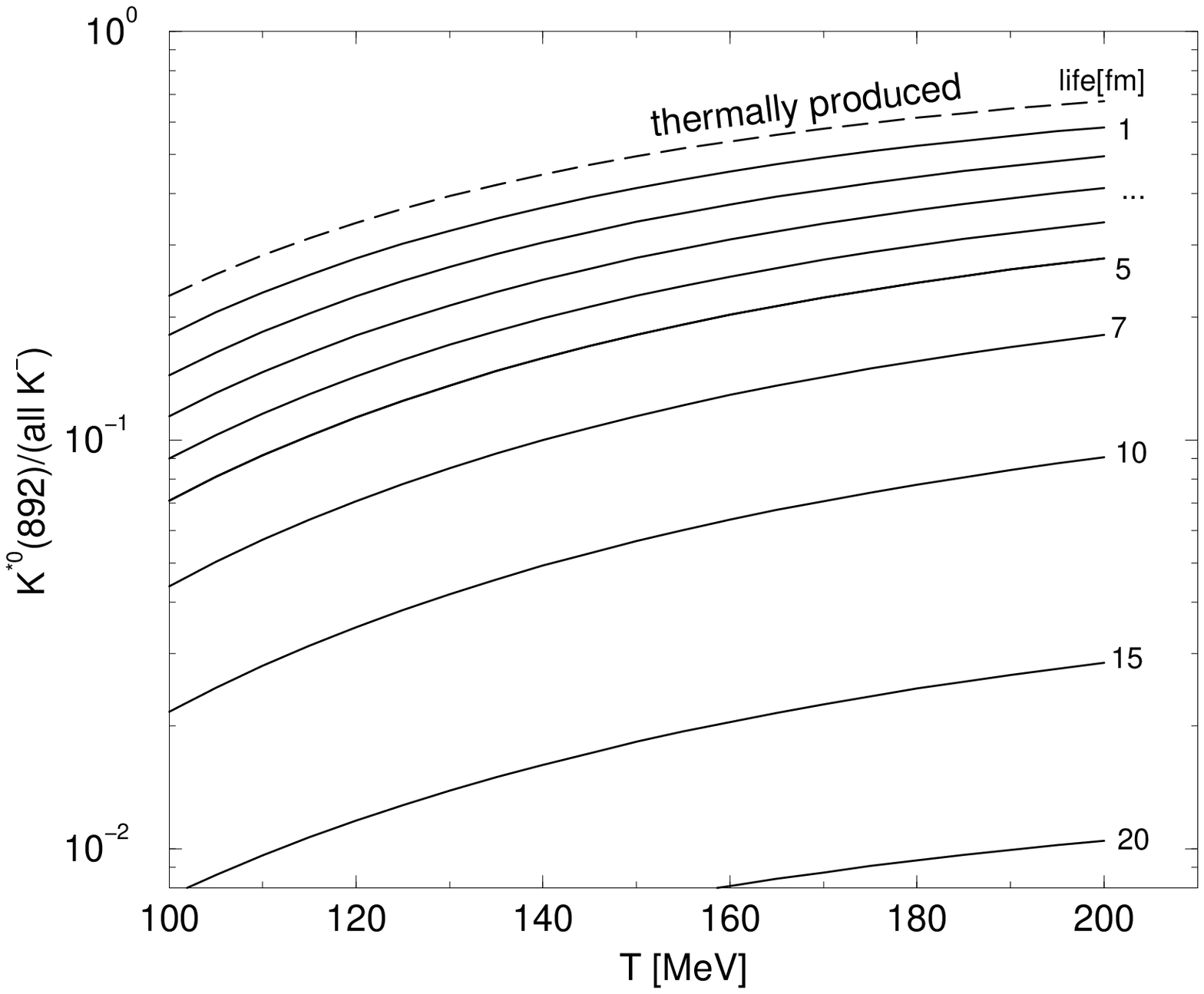}
\vspace*{0.1cm}
\caption{Produced (dashed line and observable (solid lines) ratios
$\Lambda(1520)$/(total $\Lambda$) (top) 
and $\Sigma^{*}$/(total $\Lambda$) (middle) and $K^{*0}/K$ (bottom).
The solid lines correspond to evolution after chemical freeze-out of
1,2,3,4,5,7,10,15,20 fm/c, respectively.
\label{obsratios}}
\end{center}
\vspace*{-0.9cm}
\end{figure}

The uncertain meson scattering cross sections 
(denoted in Table \ref{parameters} for $\pi\pi$ by $X$  
and for $\pi K $ by $0.5\cdot X$) are therefore not 
material for the results we present. We quantitatively explored 
the range $1<X<100$ mb without seeing any significant
change in our results. In the $\rho$ resonance region 
applicable at the conditions considered  the value $X\simeq 70$ mb
is appropriate \cite{Mar76}. Remainder of Table \ref{parameters} 
presents an overview of other   parameters used in  the model.

Figure \ref{obsratios} shows the dependence of the $\Lambda(1520)/\Lambda$,
$\Sigma^{*}/\Lambda$ and $K^{*\,0}(892)/K^+$
on the temperature and lifetime of the  interacting phase.
It is clear that, given a determination of the respective signals
to a reasonable precision, a qualitative distinction between the high temperature
chemical freeze-out scenario followed by a rescattering phase and the low
temperature sudden hadronization scenario can be made. We also note that
despite the shorter lifetime of the $\Sigma^{*}$ and
higher pion interaction cross section, 
more $\Sigma^{*}$ decay products should be reconstructible
than in the $\Lambda^{*}(1520)$ case, at all but the highest temperatures
under consideration.

Fig.~\ref{obsratios} demonstrates the sensitivity of strange hadron
 resonance production to the interaction period  in the 
hadron phase, {\it i.e.} the phase between the thermal and chemical freeze-out.
What we learn from Fig.~ \ref{obsratios} is that
while the suppression of one of these ratios considered has generally two
interpretations, as it can 
mean either a low temperature chemical freeze-out or a long interacting
phase with substantial rescattering, the comparison of
two resonances with considerably different lifetimes can be 
used to constrain both the
temperature of chemical freeze-out and the lifetime of the interacting phase.

We can reexpress the results presented in Fig.~\ref{obsratios} representing
one ratio against the other as is seen in  Fig.~\ref{projdiag}, and 
attempt a first comparison with experiment.  
Combining the $\overline{K^{*0}}/\pi+$ results 
presented in \cite{NA49Res,NA49Res2}
with earlier $K^-/\pi^-$ and $K^+/\pi^+$
results \cite{oldNA49}, it seems that the 
$\overline{K^{*0}}/K^+$ ratio should be $\approx$ 0.2.
Unless the $\Lambda(1520)/\Lambda$ ratio will be considerably 
different than what is expected from
\cite{NA49Res}, experimental measurement for SPS
energies should be in the top left corner 
of the bottom portion of the Fig.~\ref{projdiag}, corresponding
to a low chemical freeze-out temperature ($\approx 145$ MeV) and 
a short, if not negligible, time spend in the hadron re-interaction
phase.  Pending more experimental work, this evaluation serves
only to demonstrate the functioning of the method we propose.

\begin{table}
\centering
\caption{Scattering model parameters}
\begin{tabular}{|c|c|c|c|c|c|} 
$\sigma_{\pi N}$ (mb) & $\sigma_{K N}$ & $\sigma_{\pi \pi}$ &
$\sigma_{\pi K}$ &  $\sigma_{N N}$ & $\sigma_{\overline{N} N}$  \\ \hline
24 & 20 & $X$ & $0.5\cdot X$ & 24 & 50 \\ \hline
\multicolumn{2}{|c|} {$\Gamma_{\Sigma^{*}}$ } & 
\multicolumn{2}{c|} {$\Gamma_{\Lambda(1520)}$}    &
\multicolumn{2}{c|} {$\Gamma_{K^{*0}(892)}$}       \\ \hline
\multicolumn{2}{|c|} {35 MeV}& 
\multicolumn{2}{c|} {15.6 MeV} & 
\multicolumn{2}{c|} {50 MeV} \\ \hline
\multicolumn{3}{|c|} {escape rate (fm$^{-3}$)} &  
\multicolumn{3}{c|} {negligible}   \\ \hline
\multicolumn{3}{|c|} {v}          & 
\multicolumn{3}{c|} {0.5}  \\ \hline
\multicolumn{3}{|c|} {R(fm)}      & 
\multicolumn{3}{c|} {8\,145/$T$ [MeV]}   \\ \hline
\multicolumn{3}{|c|} {$\mu_b$} & 
\multicolumn{3}{c|} {$220$ MeV}   \\ \hline
\multicolumn{3}{|c|} {A} & 
\multicolumn{3}{c|} {11}   \\ \hline
\end{tabular}
\label{parameters}
\end{table}
Should the $\Sigma^{*}/\Xi$ ratio become available along with 
a determination 
of $\lambda_{i}$ and $\gamma_{i}$, both the temperature
and the lifetime can be inferred from the $\Sigma^{*}$ alone.
Fig. \ref{projdiagxi} shows an analogous diagram to the one
shown previously, but with the $\Sigma^{*}$/(all $\Lambda$) on the
horizontal axis and  $\Sigma^{*}$/(all $\Xi$) on the vertical.
Since $\Sigma^{*}/(\Xi+\Xi^{*})$ only depends on 
temperature by about $5 \%$, see Fig.\,\ref{prodratios},
the percentage of rescattered $\Sigma^{*}$  can be, to a good approximation,
read off the vertical axis in  Fig.\,\ref{projdiagxi}. 

\begin{figure}[tb]
\begin{center}
\vspace*{-1.6cm}
\hspace*{.1cm}
\psfig{width=10cm,clip=1,figure=\pathnow 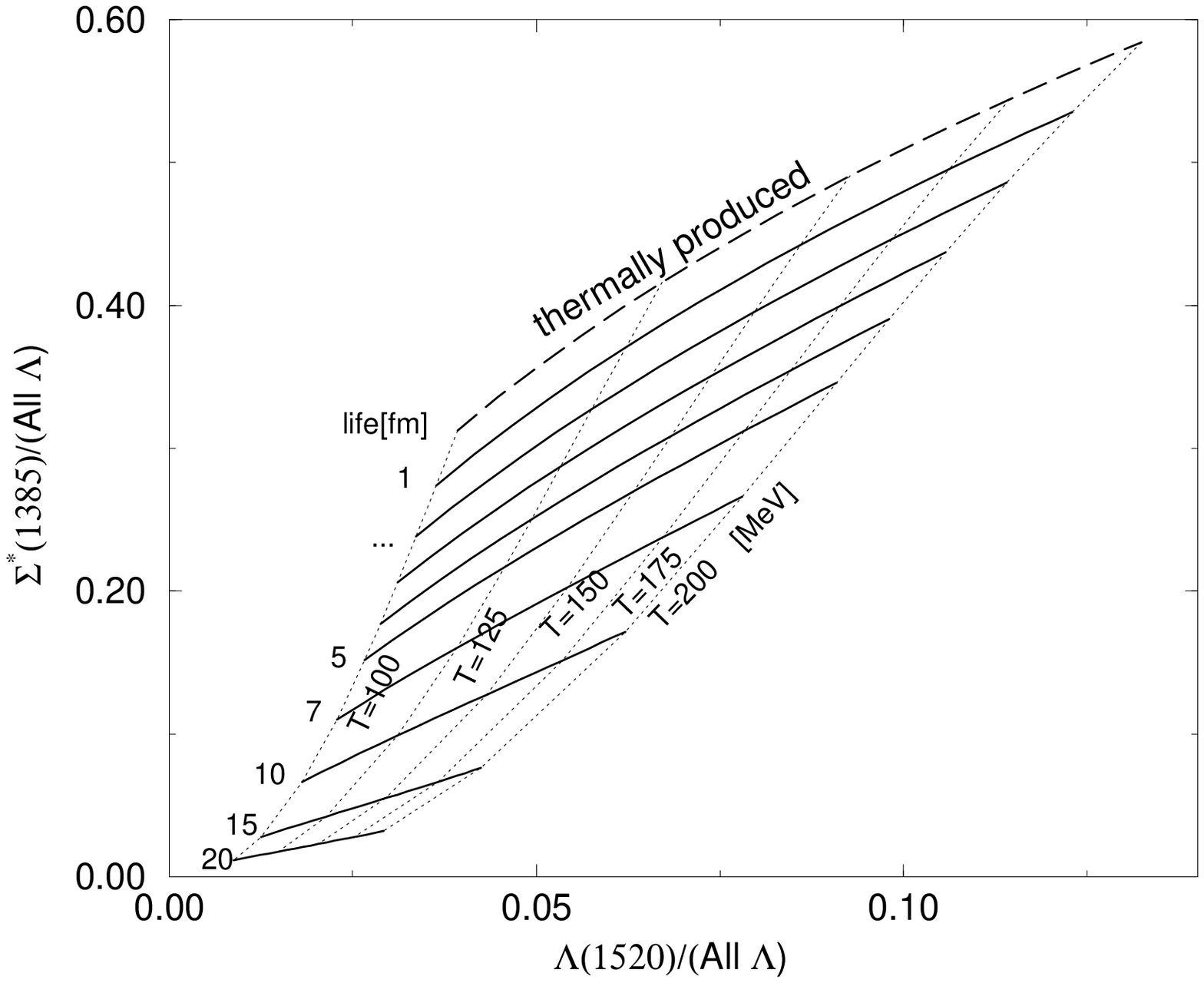 }
\vskip -1.5cm
\hspace*{.1cm}
\psfig{width=10cm,clip=1,figure=\pathnow 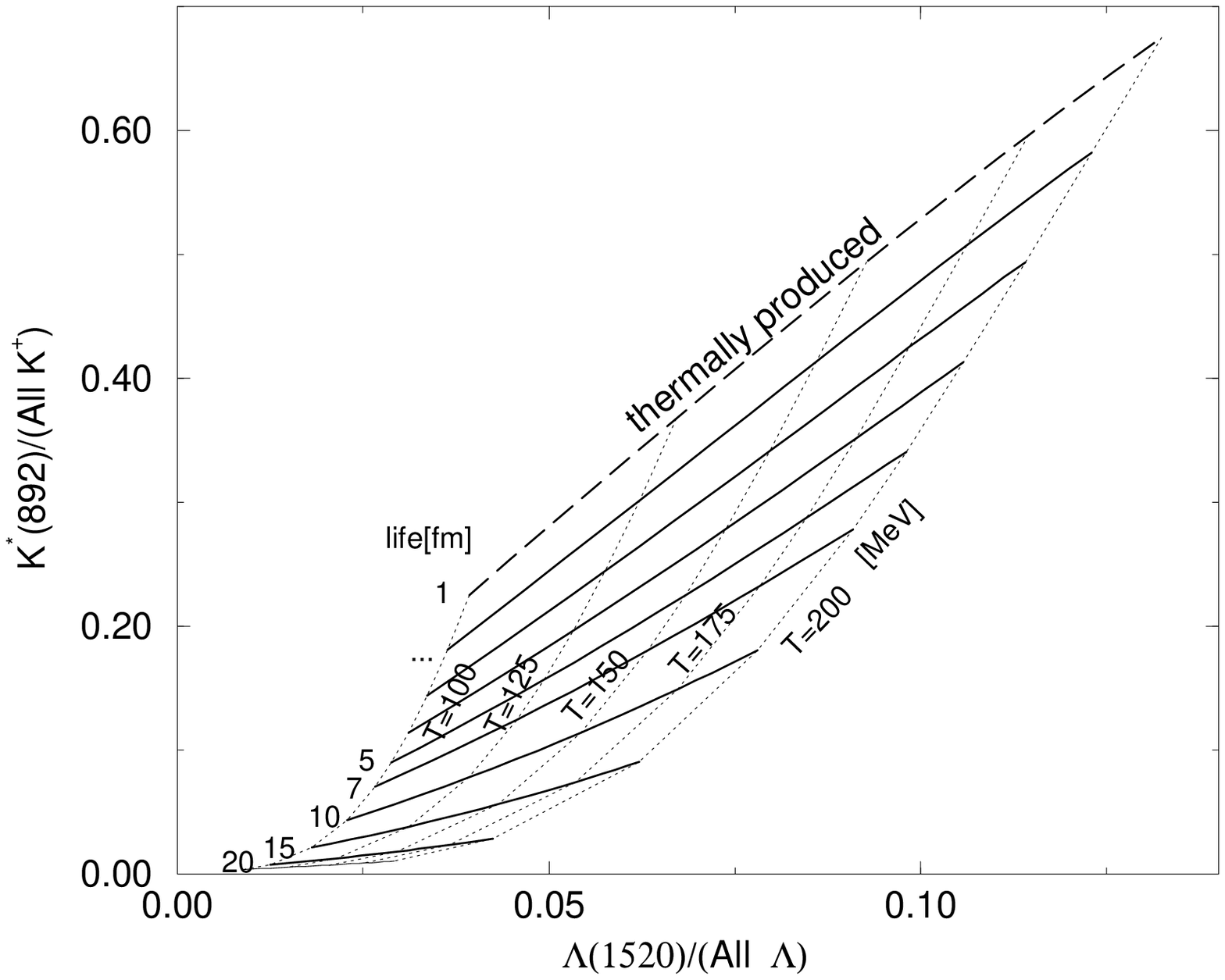}
\caption{Dependence of the combined $\Lambda(1520)$/(all $\Lambda$),
$\Sigma^{*}$/(all $\Lambda$) and $K^{*0}(892)/$(all K-)
signals on the chemical freeze-out temperature
and interacting phase lifetime.}
\label{projdiag}
\end{center}
\end{figure}
\vspace*{-.5cm}

We have shown  that a comparison of two or better three
observable strange hadron resonance abundances, with 
different masses, widths and interaction modes of
the decay products, can be used to estimate
the currently uncertain hadron freeze-out behavior.
We have chosen to consider strange hadrons as they have
relatively narrow widths.  A similar study of the 
lifespan between thermal and chemical freeze-outs, and 
chemical freeze-out temperature as parameters,
can be  undertaken  with non-strange nucleon
resonances. We believe that  the experimental 
difficulties of such a measurement are much greater:
the relative  yield of resonances such as 
$N^*(1440)$, $\Delta(1230)$ compared to the nucleon is 
smaller than in case of strange baryons, the  widths 
are much larger ($\Gamma_{N^*}\simeq 300$\,MeV, $\Gamma_{\Delta}=120$\,MeV),
and the final state nucleon has a greater scattering cross section 
in hadronic matter than the strangeness
carrying hadron. Therefore the detectability  of the non-strange
baryon resonances in terms of an invariant
mass analysis in the decay channels should be much smaller, 
except if a (very) sudden hadronization
applies.  Seen from that perspective, it will be  interesting
to see if the   $\Delta(1230)$ state can be observed at all, as this
would be suggesting that chemical and thermal freeze-out are indeed
nearly coincident.

\begin{figure}[tb]
\begin{center}
\vskip -1.8cm
\hspace*{.1cm}
\psfig{width=10cm,clip=1,figure=\pathnow 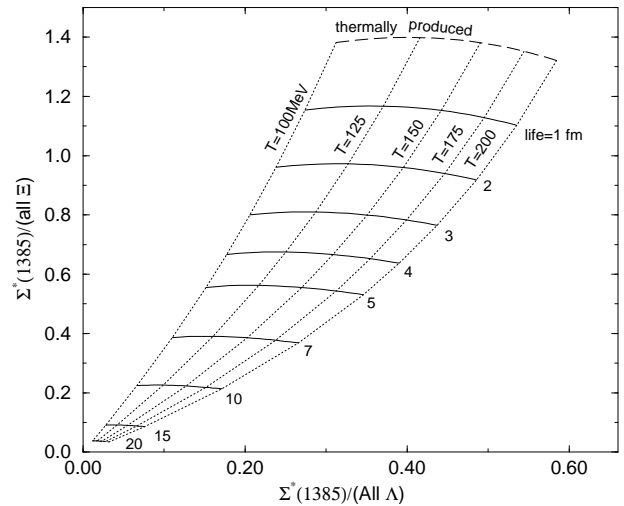 }
\vspace*{+0.1cm}
\caption{Dependence of the combined $\Sigma^{*}$/(all $\Lambda$ s) 
and
$\Sigma^{*}$/(all $\Xi$ s) signals on the chemical freeze-out temperature
and interacting phase lifetime}
\label{projdiagxi}
\end{center}
\end{figure}
\vspace*{-.5cm}

We believe that a detector capable of reconstructing $\Xi$ and $\Omega$ decays,
with a narrower invariant mass 
 resolution than the natural $\Sigma^{*}$-width 
(35\,MeV) may be capable of measuring the
$\Sigma^{*}$ signal  by performing an invariant mass fit on
$\Lambda$--$ \pi$ pairs arriving directly from the fireball. 
Present day state of the art 
detectors such as STAR and NA57 should satisfy this 
criterion.  We have shown that the $\Sigma^{*}/\Xi$ ratio
 is remarkably independent (within  5 \%) of freeze-out temperature and 
is greater than unity. A significantly depleted ratio would 
directly measure rescattering of hadrons after 
chemical freeze-out. Considering other ratios 
we learn from Fig.~\ref{obsratios}  that
while the suppression of one of particle ratios considered  is 
generally not able to allow resolution of the freeze-out properties, as it can 
mean either a low temperature chemical freeze-out or a long interacting
phase with substantial rescattering, the comparison of
two relative resonance yields with considerably different lifetimes can be 
used to constrain both the
temperature of chemical freeze-out and the lifetime of the interacting phase.
This is than qualitatively evaluated in Fig.~\ref{projdiag}.

We have shown in quantitative fashion how a measurement of the relative 
yields of $\Lambda(1520)$ and $K^{*0}(892)$ with respect to the
ground states  $\Lambda$ and $K$ allows to constrain the chemical freeze-out 
temperature, and the lifetime of the intermediate evolution
period ranging between thermal
and chemical freeze-outs in relativistic heavy ion collisions, 
and hence to distinguish between sudden and
staged hadronization scenarios. We have suggested that present 
experiments should be also able to detect
the $\Sigma^{*}$ resonance, which when subject to the same analysis 
method  offers a consistency check for  the 
reaction mechanism considered.

{\vspace{0.5cm}\noindent\it Acknowledgments:\\}
Supported  by a grant from the U.S. Department of
Energy,  DE-FG03-95ER40937\,. 




\end{narrowtext}

\end{document}